\documentclass[a4paper,12pt]{article}
\usepackage[utf8]{inputenc}
\usepackage[italian, english]{babel}
\usepackage[left=2cm,right=2cm,top=2cm,bottom=2cm]{geometry}

\usepackage{amsmath} 
\usepackage{graphicx} 
\usepackage{float}
\usepackage{amsthm} 
\usepackage{amsfonts} 
\usepackage{amssymb} 
\usepackage{xcolor} 
\usepackage{listings} 
\definecolor{backcolour}{rgb}{0.95,0.95,0.92}
\usepackage{verbatim}
\usepackage{siunitx}
\usepackage{youngtab}
\usepackage{ytableau}
\usepackage{subfig}
\usepackage{braket}
\usepackage{physics}
\usepackage{slashed}
\usepackage{booktabs}
\usepackage{tikz}
\usepackage{dynkin-diagrams}

\usepackage[T1]{fontenc} 
\usepackage{yfonts}

\usepackage[colorlinks=true, linkcolor=blue, citecolor=blue]{hyperref} 

\numberwithin{equation}{section}

\usepackage{bbm}
\usepackage[utf8]{inputenc}
\usepackage{graphicx}
\usepackage{hyphenat}
\usepackage{wrapfig}
\usepackage{empheq}
\usepackage{textcomp}
\usepackage{rotating}
\usepackage{wrapfig}
\usepackage{pdfpages}
\usepackage{verbatim}
\usepackage{setspace}
\usepackage{array}
\usepackage{caption}
\usepackage[pdf]{pstricks}
\usepackage{upgreek}
\usepackage{cmll}
\usepackage{latexsym}
\usepackage{braket}
\usepackage{cite}
\usepackage{epsfig}

\usepackage{tikz-cd}

\newcommand{\beq}{\begin{eqnarray}}
	\newcommand{\eeq}{\end{eqnarray}}
\newcommand{\bea}{\begin{eqnarray}}
	\newcommand{\eea}{\end{eqnarray}}
\newcommand{\be}{\begin{equation}}
	\newcommand{\ee}{\end{equation}}

\newcommand{\rmc}{\mathrm{c}}

\def\brc{\langle }
\def\ckt{\rangle}
\def\const{{\rm const}}

\numberwithin{equation}{section}

\begin{document}

\title{Natural Anomaly Matching}

\vskip 40pt  
\author{  
Stefano Bolognesi$^{(1,2)}$, 
 Kenichi Konishi$^{(1,2)}$, Andrea Luzio$^{(3,2)}$,\\  and  Matteo Orso$^{(1,2)}$    \\[13pt]
{\em \footnotesize
$^{(1)}$Department of Physics ``E. Fermi'', University of Pisa,}\\[-5pt]
{\em \footnotesize
Largo Pontecorvo, 3, Ed. C, 56127 Pisa, Italy}\\[2pt]
{\em \footnotesize
$^{(2)}$INFN, Sezione di Pisa,    
Largo Pontecorvo, 3, Ed. C, 56127 Pisa, Italy}\\[2pt]
{\em \footnotesize
$^{(3)}$Scuola Normale Superiore,   
Piazza dei Cavalieri, 7,  56127  Pisa, Italy}\\[2pt]
\\[1pt] 
{ \footnotesize  stefano.bolognesi@unipi.it,  kenichi.konishi@unipi.it,  andrea.luzio@sns.it,  m.orso1@studenti.unipi.it}  
} 
\date{}

\vskip 6pt

\maketitle

\begin{abstract}

In a large class of chiral gauge theories in four dimensions it was found that certain natural assumption about the 
bifermion condensates leads to the infrared effective theory where the 't Hooft anomaly matching conditions are satisfied in an entirely evident fashion, without need of verifying arithmetic equations. This is due to the fact that  in these systems, characterized by dynamical color (and  flavor) symmetry breaking, the quantum numbers and multiplicities of the low-energy massless fermions match exactly those  of the fermions in the UV theory which  do not condense and remain massless,  with respect to the unbroken symmetries.  This means also that the stronger constraints following from the matching request of 
 mixed anomalies involving certain generalized 1-form center symmetries, as well as some global anomalies such as 
Witten's  $SU(2)$ anomaly,  are all fully satisfied. 
  It is the aim of this note to clarify better the working of this phenomenon (which we call  Natural Anomaly Matching), 
  correct some earlier statements made about it,  and illustrate it further with a  few  new examples.

\end{abstract}

\maketitle
	
	\newpage

	\tableofcontents

	\newpage 
		
	\section{Introduction}

	After many years of investigations \cite{tHooft}-\cite{ShifmanUnsal},  characteristics of the dynamics of strongly coupled (especially, chiral) gauge theories in four dimensions are gradually emerging \cite{BKS}-\cite{SheuShif}.  One of the key ideas which made the recent progress possible is that of the generalized symmetries, and the possibilities of gauging some higher-fom symmetries and the resulting new, mixed types of anomalies \cite{KS}-\cite{AnberChan}.  Consistency requires that such new anomalies be reproduced in the infrared appropriately, a constraint which is in general  stronger than the one following from the conventional 't Hooft anomaly matching arguments \cite{tHooft}.

     Independently of, and consistently with, these exciting new developments,  it was noted in our previous work \cite{BK,BKL2,BKL4,BKLDA},  that  the  anomaly-matching condition is sometimes satisfied totally evidently,  i.e., without the need for solving any arithmetic  matching equations among the anomaly coefficients. 
This occurs when the fermions participating in the (bi-fermion) condensates get mass dynamically, forming massive Dirac pairs, while the remaining massless fermions are identical, both in the quantum numbers and multiplicities, with respect to the unbroken symmetries,  with a simple set of weakly-coupled massless fermions in the infrared.

 In the color-flavor-locked dynamical Higgs phase \cite{BK, BKL2, BKL4},  the latter  correspond to gauge-singlet composite fermions (``baryons'').  In systems with dynamical Abelianization \cite{BKS,BKLDA}, instead, the low-energy massless fermions are simply the subset of the original fermions, which do not participate in the condensates and  which become weakly coupled below the condensation mass scale.

In both cases, the set of massless fermions in the UV and IR have the same quantum numbers and multiplicities 
with respect to the surviving symmetries.  Thus 
 the anomaly-matching requirement is satisfied both for the conventional, perturbative  't Hooft anomalies and  for the    
generalized anomalies which take into account the global structure of the symmetry groups and the possible gauging of certain  
1-form center symmetries. Witten's $SU(2)$ anomaly is also guaranteed to be preserved.  

The phenomenon (which we may call Natural Anomaly Matching) looks simple and reasonable; nevertheless  it was reported in 
 \cite{BK}  where a large class of chiral gauge theories have been surveyed,  that in some cases the natural anomaly matching seemed to fail. 
 
 It is the aim of this work to clarify better the working of natural anomaly matching, discuss some of its characteristics,  and illustrate it  further with some new examples in the context of strongly-coupled tensorial chiral gauge theories.
 
 
In Sec.~\ref{NAM}  we discuss the essential features of the natural anomaly matching.  In Sec.~\ref{CF} the cases of color-flavor-locked dynamical Higgs phase are illustrated by first reviewing a few well known examples, and then discussing some new examples. A few systems which might reduce dynamically to infrared-free effective gauge theories, either fully Abelian or nonAbelian, are explored in Sec.~\ref{IRfree}. Summary and conclusion are in Sec.~\ref{Concl}.

\section{Natural Anomaly Matching \label{NAM} }

Let us indicate the original (lefthanded) fermions as a set  $\{\Psi\}$.  Pairwise interactions among  the members of $\{\Psi\}$  get strong in the infrared due to the color $G_c$  gauge interactions.  Depending on the representation of $G_c$ to which  $\{\Psi\}$  belong, 
the system  has a global flavor symmetry group, $G_f$.
 Let us divide the full set $\{\Psi\}$  as a sum
\be     \{\Psi\}  =    \{\Psi_0 \}    \oplus    \{\Psi_1\}  \;, 
\ee
where the first group of fermions form pairwise condensation  in the vacuum 
\be   \brc  {\Psi_0} {\Psi_0}  \ckt \sim   \Lambda^3  \;,  \label{condens} 
\ee

where $\Lambda$ is of the order of the renormalization-group invariant mass scale of the $G_c$ gauge theory.  The vacuum expectation values (VEV) (\ref{condens})  break the 
color and flavor symmetries as 
\be     G_c \times  G_f  \to    {\tilde G}_c  \times     {\tilde G}_f\;,   \label{IRsym} 
\ee
where 
\be      {\tilde G}_c \subset G_c\;, \qquad    {\tilde G}_f \subset   (G_c/  {\tilde G}_c )   \times  G_f\;, \label{allow}
\ee
and $ {\tilde G}_c$  indicate the surviving gauge group in the infrared.  Namely, the low-energy theory is a $ {\tilde G}_c$  
gauge theory with fermions $ \{\Psi_1\}$ in various representations of ${\tilde G}_f.$

In (\ref{allow}) we have taken into account the possibility that the condensates (\ref{condens})  break both (part of) color and flavor 
symmetry groups, but leave some diagonal combination invariant, i.e.,  the color-flavor locking.

When the  surviving color gauge interactions with $ {\tilde G}_c$ are asymptotically free,  the system further evolves towards lower energies  (this is known as 
``tumbling'' \cite{RabyDim})  and the fate of the symmetries depend on the dynamics of  $ {\tilde G}_c$ interactions with fermions  $\{\Psi_1\}$.
 As the problem is essentially of the same nature as the original one, we shall assume from here on that 
$ {\tilde G}_c$   is either trivial  or infrared-free \footnote{
Our main interest is in chiral gauge theories.  In vectorlike theory our construction has not much to add to what is known about their dynamics.  For instance, in the standard QCD,  $\Psi_0 =  \{ \psi_L^i,  (\psi_R^c)^{i}\},$   $i=1,2, \ldots, N_f$,   $\{\Psi_1\} = \emptyset$.   $G_c=SU(3),$ $ {\tilde G}_c=\emptyset$,  
$G_f=SU(N_f)\times SU(N_f)\times U_V(1)\times U_A(1)$,   ${\tilde G}_f =   SU(N_f)_V\times U_V(1)$.  There are no massless fermions in the IR but the anomaly matching does not pose a problem as there are no anomalies associated with  ${\tilde G}_f$ in the UV.  All anomalous symmetries are spontaneously broken.}.

Now each pair  of fermions going into the condensates $ {\Psi_0} {\Psi_0}$ form together a singlet of  ${\tilde G}_f$. In other words  they do not contribute to the  ${\tilde G}_f^3 $    anomalies \footnote{Below we indicate by  a symbol  $\Tr  T^3$ the symmetric trace of three generic generators  $\Tr \left( \{T^a T^b\} T^c \right)$  of  a given algebra. }
\be  {\Tr} \,  {\tilde T}^3|_{\Psi_0} =0\;, \qquad      {\tilde T}  \subset   \{  T_c + T_f \}\;.  \label{used}  \ee
Next  observe that a simply relation 
\be   {\Tr} \,    T_f^3  |_{{\Psi_0}+ {\Psi_1}}=  {\Tr} \,   (T_c+ T_f)^3  |_{{\Psi_0}+ {\Psi_1}}     \;\label{simple} 
\ee
holds for {\it  any}  generators  ($T_c, T_f)$.  To see this,  use the fact that $G_c$ is nonanomalous with the fermions $ \{\Psi_0, \Psi_1\} $,  and  that $G_f$ 
represent a nonanomalous flavor symmetry.   Also all generators of $G_c=SU(N)$ are traceless. That is, 
\be 
\Tr\left[T_f^2 T_c\right]|_{{\Psi_0}+ {\Psi_1}}  =  \Tr\left[T_c^2 T_f\right]|_{{\Psi_0}+ {\Psi_1}} =\Tr\left[T_c^3\right]|_{{\Psi_0}+ {\Psi_1}} =0\,. \ee
Let us now apply  (\ref{simple})  to the generators $T_c$  and  $T_f$ which enter  $ {\tilde T}$.  It follows that 
\be     \Tr    T_f^3  |_{{\Psi_0}+ {\Psi_1}}= \Tr   (T_c+ T_f)^3  |_{{\Psi_0}+ {\Psi_1}}=    {\Tr}\,   {\tilde T}^3|_{{\Psi_0}+ {\Psi_1}}=
 {\Tr}\,   {\tilde T}^3|_{\Psi_1}\;,  \label{This}
\ee
where (\ref{used}) was used at the final step.

This looks like the solution of the UV-IR anomaly matching condition.
Actually, this does not in itself explain why in some chiral gauge theories we find a natural set of massless composite fermions 
(baryons) saturating all the anomalies, in an obvious fashion.  
The problem is that at the scale $\Lambda$ the fermions $\{\Psi_1\}$ are strongly coupled, 
therefore in order to have the  ${\tilde G}_f$ symmetry properly realized at low energies, 
 all fermions in the set $\{\Psi_1\}$ must be replaced by  a set of $G_c$ invariant ``baryons''  $\{{\cal B}\} $ which are weakly coupled at
$\mu \lesssim  \Lambda$  and which reproduce all the  ${\tilde G}_f$  anomalies  due to $\{\Psi_1\}$, i.e., 
\be    {\Tr}\,   {\tilde T}^3|_{\Psi_1}   =   {\Tr}\,   {\tilde T}^3|_{\cal B}\;.  \label{theproblem}  
\ee
 But then, a priori,  this looks just the original 't Hooft's
problem:  there is no guarantee in general that an appropriate set of baryons saturating  (\ref{theproblem})  exists or  can be found.

What happens in certain models with a color-flavor-locked dynamical Higgs vacuum \cite{ADS,BK,BKL2,BKL4}  is that the VEV of the composite scalar 
$\Phi\sim  \Psi_0 \Psi_0$  acts  as a metric, which can be used to convert the color to the flavor indices. In the simplest cases where 
\be    \brc  \Phi^i_a \ckt  =   \brc  {\Psi_0} {\Psi_0}  \ckt^i_a  \sim   \delta^i_a   \Lambda^3\;,   \label{CFlocking} 
\ee 
where  $i$ is the color index in the fundamental representation, ${\underline N}$,  of $G_c=SU(N)$, and $a$ is an index of the flavor 
symmetry $G_f$,   all color indices in  $\Psi_1$  can be eliminated (i.e., made gauge invariant) and  converted to flavor labels, in the ``baryons'' of the form,   
\be    \{{\cal B}\} \sim  \Psi_1 \Psi_0 \Psi_0   \;,  \quad   \Psi_1   \Psi_0^* \Psi_0^*   \;,  \quad   \Psi_1 \Psi_0 \Psi_0   \Psi_0 \Psi_0 \;,  
\quad   \Psi_1 \Psi_0 \Psi_0  \Psi_0^* \Psi_0^*   \;,  \quad   \Psi_1  \Psi_0^* \Psi_0^*  \,  \Psi_0^* \Psi_0^*  \;,  \label{baryons}   
\ee
and so on.  Which massless baryons among (\ref{baryons})  actually form, depend on the $G_c$ representation each of  $\Psi_1$ belongs to.  Once an appropriate set of the baryons are found, which replace in the infrared  all the UV fermions  $\Psi_1$,   
all the issues about perturbative and nonperturbative anomaly matching literally disappear:   see Sec.~\ref{CF} 
 below for examples. 


A second possible situation in which the problem (\ref{theproblem}) has a simple  solution is when the low-energy color gauge group in (\ref{IRsym})  is infrared free. In those cases,  the remaining massless fermions $\{\Psi_1\}$ survive in the IR, become weakly coupled, and carry all the information about the unbroken symmetry group ${\tilde G}_f$ into the infrared.  The full anomaly matching, (\ref{This}),
 is again automatic.   A well studied case is that of a ``$\psi\chi\eta$'' model,  where the system is likely to dynamically Abelianize \cite{BKS,BKLDA}, somewhat analogously to what happens in many ${\cal N}=2$ supersymmetric gauge theories, where
\be       {\tilde G}_c = \prod_{i=1}^{N-1}  U_i(1)  \subset   G_c= SU(N)\;. 
\ee 
 In   Sec.~\ref{IRfree}  we briefly review these cases, and discuss some new situation where $  {\tilde G}_c $  is infrared free but  nonAbelian. 


The ideas related to the Natural Anomaly Matching appeared earlier in various forms \cite{RabyDim}-\cite{ShifmanUnsal};  
they may be ultimately traced back to the original work on tumbling and complementarity, in the context of dynamical Higgs
vacua \cite{RabyDim}. The present note is an attempt to streamline the logic under them, discuss its systematic features,
the range and limit of their applications, in the light of the recent development involving the generalized symmetries and stronger 
form of 't Hooft anomaly-matching discussions.   
In fact, one of the main outcomes of the  recent studies of a wide range of chiral gauge theories in four dimensions \cite{BKL1}-\cite{corfu}, 
is that certain ``non-anomalous''  discrete or $U(1)$ symmetries become anomalous under the  gauging of the $1$-form  color ${\mathbbm Z}_N \subset SU(N)$  center symmetry.  And this, in all cases studied,  is consistent with the assumption of a dynamical Higgs
 phase or dynamical Abelianization  \cite{BKLDA,SheuShif},  induced by some bifermion condensates.  Then, under a reasonable choice of the condensates,  the question of anomaly-matching (consistency check) with respect to the unbroken symmetries  becomes of a simple resolution:  the conventional, new and global anomalies are all UV-IR matched,  manifestly.

\subsection{``Non-gauge-invariant condensates''}  

%
 In chiral gauge theories, a bifermion condensate (\ref{condens}) is in general gauge-dependent. Such a condensate must eventually be understood as the VEV of a composite gauge-invariant field, implying a nonvanishing value for some of its components in a chosen gauge. This fact does not however mean that (\ref{condens}) is either unphysical or to be excluded \`a priori. 
Quite the contrary. A gauge-noninvariant VEV (hence colored Nambu-Goldstone bosons, to be absorbed by the longitudinal components of the gauge bosons) is the essence of the Higgs mechanism. As is well known, this does not mean spontaneous breaking of a local gauge symmetry.   The latter is not really a symmetry:  it cannot be broken spontaneously \footnote{Indeed, Elitsur's theorem \cite{Elitsur} should not be understood as something that excludes the Higgs mechanism.  }.

%

Usually,  the Higgs mechanism is taught  in textbooks in the context of a weakly coupled gauge theory, with a simple potential for the Higgs scalar field with nontrivial minima, i.e., its nonvanishing VEV.  In the strongly-coupled chiral gauge theories we are concerned with here,  the scalar field emerges as a bifermion composite  
by the strong gauge interactions, and is itself, in general, strongly coupled.  Nevertheless, there is no essential difference from the standard Higgs mechanism (hence the name of dynamical Higgs mechanism \cite{RabyDim}), see \cite{BKL1,BKL5} for some more discussions.

Another important aspect of a non-gauge-invariant VEV such as (\ref{condens}), which does not seem to  be appreciated  generally,  is the following. Even though the condensate (\ref{condens}) depends on the gauge choice,   
its implication on the flavor symmetry breaking is a genuine, physical one.  For a familiar example of  this, consider the standard (minimal) Glashow-Weinberg-Salam electroweak theory with a quartic Higgs scalar potential. An accidental would-be global $SO(4)$ symmetry of the Higgs Lagrangian, present in the limit of vanishing  $SU(2) \times U(1)$ gauge coupling, is broken spontaneously to $SO(3)$   by the Higgs scalar VEV 
\be      \brc    {\cal R} \phi  \ckt =    \left(\begin{array}{c}  v \\0\end{array}\right)\;, \qquad   v   \ne 0\;,    \label{WS2}
\ee
(${\cal R}$ standing for the real part).  
     The $W$ and $Z$ boson masses following from the Higgs kinetic term satisfy the relation 
\be   \rho \equiv \frac{M_{W}^2}{M_Z^2 \cos^2 \theta_{W}} = 1 + O(\alpha)\;,  \label{custodial} 
\ee
where $O(\alpha)$ indicates generic corrections of the order of the $SU(2) \times U(1)$ gauge coupling constant squared. 
(\ref{custodial}) is a physical prediction of the minimal standard model  with the Higgs VEV  (\ref{WS2}), well met by the experiments.    As is well known,   this relation is 
a manifestation of the $SO(3) \subset SO(4)$ symmetry  (known as the custodial $SU(2)$ symmetry) left unbroken by the condensate, (\ref{WS2}).    

Another example is the Bars-Yankielowicz  (BY) model  reviewed below, in  Sec.~\ref{sec:BY}.  The bifermion VEV  (\ref{cflocking}) is a gauge-dependent expression,  but the reduction of the flavor symmetry
\be   SU(N+4) \times {U}(1)  \to SU(N) \times  SU(4)  \times U(1)\,  \ee 
implied by it,  is a gauge-independent conclusion about the physics of this model.

The lesson to be drawn is that even though a gauge-non-invariant VEV (elementary or composite), commonly used in the literature, does not mean spontaneous breaking of a gauge symmetry,  it can imply spontaneous breaking of a {\it  flavor} symmetry.

\section{Color-Flavor Locking and dynamical Higgs phase   \label{CF}  }

In this section we illustrate and discuss  the idea of color-flavor-locking  and natural anomaly matching with some examples.
  In particular,  
chiral $SU(N)$ gauge theories  with matter chiral fermions in the representation, 
\be \Yvcentermath1      N_{\psi}  \, \yng(2) \oplus     N_{\chi}\,    {\bar  {\yng(1,1)} }   \oplus    n  \,    {\bar  {\yng(1)}} \;, \label{classification}  
\ee
where 
\be    n=  N_{\psi}  (N+4) -    N_{\chi} (N-4)  \;, 
\ee 
 have been studied in  \cite{BK},  for  $(N_{\psi},   N_{\chi})=
(1,0), (2,0), (3,0),(0,1),(0,2),(0,3),(2,1),(1,1),$ $(1,-1)$.   We review a few well-known cases,  correct some statements made in \cite{BK}, and
add a few more examples.

Though we are not going to review here in detail,  the  idea of color-flavor locking and  dynamical Higgs phase for these classes of chiral gauge theories have recently  found a theoretical support from the analyses of the  global structure of the symmetry group, and gauging  of the 
1-form center symmetries and new, mixed type anomalies \cite{BKL1}-\cite{BKLReview}.

\subsection{Bars-Yankielowicz  (BY) model  \label{sec:BY} } 

The best known example is the BY model ($\psi\eta$ model) discussed in \cite{BY}-\cite{ADS}, \cite{BKL2}.  
The matter fermions are
\beq
   \psi^{\{ij\}}\,, \quad    \eta_i^B\, , \qquad \;  B=1,2,\ldots , N+4\,,
\eeq
or
\be
\Yvcentermath1
       \yng(2) \oplus   (N+4) \, {\bar {\yng(1)}}\;. 
\ee
The (continuous) symmetry of this model is 
\beq
SU(N)_{\rm c} \times  SU(N+4)_{\rm f} \times {U}(1)\,,  \label{full}
    \eeq
    where $U(1)$ is an anomaly-free combination of  $U_{\psi}(1)$ and $U_{\eta}(1)$, with 
    \be     Q_{\psi}:\quad   N+4 \;, \qquad  Q_{\eta}:\quad    -(N+2) \;.  \label{comb}
    \ee
A color-flavor locked dynamical Higgs phase \cite{ADS,BKS,BKL2}, is characterized by  
\be    \brc  \psi^{\{ij\}}   \eta_i^B \ckt =\,   c \,  \Lambda^3   \delta^{j B}\;,   \qquad   j, B=1,2,\dots  N\;,    \label{cflocking}
\ee
in which the symmetry is  reduced to 
\beq
SU(N)_{\rm cf} \times  SU(4)_{\rm f}  \times U^{\prime}(1) \,.
    \eeq
    The candidate massless baryons are:
      \be     B^{[A B]}=    \psi^{ij}  \eta_i^{A}  \eta_j^{B}  \;,\qquad  A,B=1,2, \ldots, N\;,
\ee
antisymmetric in  $A \leftrightarrow B$, and 
  \be  B^{[A C]}=    \psi^{ij}  \eta_i^{A}  \eta_j^{C}  \;,\qquad  A=1,2, \ldots, N\;,\quad  C=N+1, \ldots, N+4\;. 
\ee
See Table~\ref{Simplest}.

\begin{table}[h!t]
  \centering 
  \begin{tabular}{|c|c|c |c|c|  }
\hline
$ \phantom{{{   {  { \yng(1)}}}}}\!  \! \! \! \! \!\!\!$   & fields   &  $SU(N)_{\rm cf}  $    &  $ SU(4)_{\rm f}$     &   $  U^{\prime}(1)   $  \\
 \hline
   \phantom{\huge i}$ \! \!\!\!\!$  {\rm UV}&  $\psi$   &   $ \Yvcentermath1 { \yng(2)} $  &    $  \frac{N(N+1)}{2} \cdot   (\cdot) $    & $   1  $    \\[1.5ex]
 & $ \eta^{A}_i$      &   $ \Yvcentermath1  {\bar  {\yng(2)}} \oplus {\bar  {\yng(1,1)}}  $     & $N^2 \cdot (\cdot )  $     &   $ - 1 $ \\[1.5ex]
&  $ \eta^{C}_i$      &   $ 4   \cdot \Yvcentermath1  {\bar  {\yng(1)}}   $     & $N \cdot  \Yvcentermath1 {\yng(1)}  $     &   $ - \frac{1}{2}  $ \\[1.5ex]
   \hline 
   $ \phantom{{\bar{ \bar  {\bar  {\yng(1,1)}}}}}\!  \! \!\! \! \!  \!\!\!$  {\rm IR}&      $ B^{[A  B]}$      &  $ {\bar  {\yng(1,1)}}   $         &  $  \frac{N(N-1)}{2} \cdot  (\cdot) $        &    $   -1 $   \\[1.5ex]
       &   $ B^{[A C ]}$      &  $   4 \cdot \Yvcentermath1{\bar  {\yng(1)}}   $         &  $N \cdot \Yvcentermath1 {\yng(1)}  $        &    $ - \frac{1}{2}$   \\[1.5ex]
\hline
\end{tabular}  
  \caption{\footnotesize   Color-flavor locked phase in the BY  model.
  $A$ or $B$  stand for  $A,B=1,2,\ldots, N$,   $C$   the rest of the flavor indices.  After the pair, ($\psi$ and the  symmetric part of  $ \eta^{A}_i$), form a massive Dirac particle and decouple,  the rest of the fermions in the UV and in the IR are identical.  
   }\label{Simplest}
\end{table}

\subsection{Georgi-Glashow  ($(N_{\psi},N_{\chi}) = (0,1)$)   model} 

In the case of the  Georgi-Glashow  model studied in \cite{RabyDim,ACSS,ADS,Poppitz,Shrock,BKS}, the matter fermions are 
 \beq
   \qquad   \chi_{[ij]}  \;, \qquad {\tilde  \eta}^{B\, j} \;,  \qquad   \quad  B=1,2,\ldots, (N-4) \,,
\eeq
or
\be \Yvcentermath1   {\bar  {\yng(1,1)} }   \oplus    (N-4)\,    \yng(1)\;. 
\ee
The  (continuous) symmetry is
\be     SU(N)_{\rm c}  \times  SU(N-4)_{\rm f} \times U(1)\;,
\ee
where  the anomaly free $U(1)$ charge  is 
\be   \chi:    \quad    N-4  \;; \qquad {\tilde  \eta}^{B\, j} \;:  \quad   - (N-2)  \;. 
\ee
%
    It was pointed out \cite{ADS} that this system may develop a condensate of the form
\be   \brc   \chi_{[ij]} {\tilde  \eta}^{B\, j}   \ckt  = \const   \, \Lambda^3 \delta_i^B \;, \qquad i, B=1,2,\ldots, N-4\;.  \label{cfl01}
\ee
The symmetry is broken to 
\beq
   SU(N-4)_{\rm cf}  \times  U^{'}(1)   \times   SU(4)_{\rm c}\,.
    \label{b270}
\eeq

Assume that the massless baryons are:
\be      B^{\{A B\}} = \chi_{[ij]} \,  \tilde{\eta}^{i\, A}   \tilde{\eta}^{j\, B}   \;, \qquad   A,B=1,2,\ldots (N-4)\;;
\ee
 the saturation of all of the anomalies associated with the remaining flavor group  $ SU(N-4)_{\rm cf}  \times  U^{'}(1)$
 is manifest  in Table~\ref{SimplestAgain2}.  
\begin{table}[h!t]
  \centering 
  \begin{tabular}{|c|c|c |c|c|  }
\hline
$ \phantom{{{   {  {\yng(1)}}}}}\!  \! \! \! \! \!\!\!$   & fields     &  $ SU(N-4)_{\rm cf} $     &   $ U^{\prime}(1) $     &  $SU(4)_{\rm c}  $     \\
 \hline
   $ \phantom{{\bar{ \bar  {\bar  {\yng(1,1)}}}}}\!  \! \!\! \! \!  \!\!\!$  {\rm UV}&  $\chi_{i_1 j_1}$     &    $\Yvcentermath1  {  \bar  { \yng(1,1)}}   $    & $N$   &   $\frac{(N-4)(N-5)}{2}\cdot (\cdot)  $ \\[1.7ex]
 &  $\chi_{i_1 j_2}$   &    $  4   \cdot \Yvcentermath1  {\bar  { \yng(1)}} $    & $\frac{N}{2}$   &   $ (N-4) \cdot  \Yvcentermath1 {\bar  { \yng(1)}}   $     \\[1.5ex]
 &$\chi_{i_2 j_2}$   &    $  \frac{4 \cdot 3}{2} \cdot (\cdot) $    & $0$    &   $ \Yvcentermath1 {\bar   { \yng(1,1)}}   $     \\[1.5ex]
& $ {\tilde \eta}^{A, \,i_1}$          & $\Yvcentermath1  \yng(2) \oplus \yng(1,1)$     &   $ - N $    &   $  (N-4)^2  \cdot  (\cdot)   $  \\[1.5ex]
 & $ {\tilde \eta}^{A, \,i_2}$         & $4\cdot \Yvcentermath1 {\yng (1)}  $     &   $ - \frac{N}{2} $     &   $  (N-4)  \cdot \Yvcentermath1  \yng(1)  $  \\[1.5ex]
   \hline 
     \phantom{\huge i}$ \! \!\!\!\!$  {\rm IR}&     $ B^{\{AB\}}$        &  $ \Yvcentermath1 {\yng(2)}$        &    $ - N $     &  $  \frac{(N-4)(N-3)}{2} \cdot ( \cdot )    $      \\[1.5ex]
\hline
\end{tabular}
  \caption{\footnotesize  Color-flavor locking  in the $(0,1)$ model.    The color index $i_1$ or $j_1$  runs up to $N-4$ and the rest is indicated by $i_2$ or $j_2$.}\label{SimplestAgain2}
\end{table}

It is interesting to note a few new  features here with respect to the BY model discussed in the previous section. The first is that the color-flavor locking here involves only a subgroup of  color $SU(N)$, due to the fact that the flavor group  $SU(N-4)$  is smaller
than the color gauge group.  Only the subgroup $SU(N-4) \subset SU(N)$ of the strong gauge group  \footnote{We assume here that   $N \ge  6$. }   can be locked with the flavor symmetry. 
The pairs   $\{ \chi_{i_1 j_1},  {\rm antisymm. \, part \, of } \,\, {\tilde \eta}^{A, \,i_1}\}$   as well as the pair   $ \{  \chi_{i_1 j_2}, {\tilde \eta}^{A, \,i_2} \}$  condense and become massive Dirac fermions.   The  $SU(4)_c\subset SU(N)$ gauge interactions remain asymptotically free. The only massless  fermions $\chi_{i_2 j_2}$ charged in $SU(4)_c$, being self-adjoint and strongly coupled,   can condense and get massive on its own. The massive ``meson'' made of  $\chi_{i_2 j_2}$ pair can be thought of as a toy model version of the dark matter,  very weakly coupled to the ``visible sector'', made of the massless 
fermions    $ B^{\{AB\}}$ of the low-energy effective theory.

Many other examples of the color-flavor locking and consequent Natural Anomaly Matching  (identical sets of massless fermions in the UV and IR, with respect to all the surviving symmetries) have been discussed in  \cite{BK},  see Tables given there, for the models $(N_{\psi},   N_{\chi})= (1,0), (2,0), (3,0),(0,1),(0,2),$  $(0,3),(2,1),(1,1), (1,-1)$. 

In a few cases, it was noted that the natural anomaly matching appeared to fail. In order to point out the reason why it did, let us consider a 
particular model  (the   $(2,1)$  model) below, and see how the introduction of an appropriate set of baryons allowed by (\ref{baryons}) restores the natural anomaly matching.  An analogous consideration allows to eliminate all the difficulties found (and some errors made)  in  \cite{BK}.

\subsection{An example  with the  $(N_{\psi},N_{\chi}) = (2,1)$ model  }
\label{nove}

%

One of the models in which the natural anomaly matching seemed to fail was the $(2,1)$  model, an $SU(N)$ theory  with fermions,   
\be 
   \psi^{\{ij\},\, m}\;, \qquad \chi_{[ij]} \;, \qquad  \eta^B_j\;,  \qquad    m=1,2,\,\quad  B=1,2,\ldots, N + 12\,,
\ee
or
\be \Yvcentermath1    2\,  \yng(2) \oplus  {\bar {\yng(1,1)}} \oplus   (N+12) \, {\bar {\yng(1)}}\;. 
\ee
The symmetries of the theory are
\beq
SU(N)_{\rm c}  \times SU(2)_{\psi} \times SU(N+12)_{\chi} \times U(1)^2\;. 
\eeq
The two $U(1)$'s are anomaly-free combinations of $U_{\psi}(1)$, $U_{\chi}(1)$, $U_{\eta}(1)$, 
which  can be taken as 
\bea &&   U_1(1): \qquad   \psi \to  e^{i  \frac{\alpha}{2(N+2)}} \psi\;, \qquad \eta \to   e^{ -i \frac{\alpha}{N+12}} \eta\;;   
\nonumber \\
&&   U_2(1): \qquad   \psi\to  e^{i  \frac{\beta}{2(N+2)}} \psi\;, \qquad \chi \to   e^{- i \frac{\beta}{N-2}} \chi\;. \label{nonanU12}
\eea
%
%
 A possibility is that a  (partial) color-flavor locked condensate  develops \cite{BK}: 
\be   \langle  \psi^{\{ij\},\, 1} \eta_{j}^B \rangle   = \, c \,  \Lambda^3 \delta^{i B} \;,   \qquad  i, B=1,2,\ldots, N  
\label{singleCF} 
\ee
where the direction of the $SU(2)_{\psi}$ breaking is  arbitrary.  Let us assume that $ \langle  \psi \chi \rangle =0\,.$

The IR symmetry of the model is 
\be
 SU(N)_{\rm cf} \times   SU(12)_{\rm f}  \times   {\tilde U}(1) \times \hat{U}(1) \times \check{U}(1) \,,  \label{no2bis}
\ee
where  $ {\tilde U}(1)$ is the unbroken combination of $U_1(1)$ and $U_2(1)$  with charges 
\be    Q_{\psi}=  1\;,   \qquad  Q_{\chi}=  - \frac{N-8}{N-2}  \;, \qquad    Q_{\eta}=  -1 \ .
\ee
$\hat{U}(1)$ and $\check{U}(1)$ are generated respectively by
\be
\hat{Q}= Q_{\psi\eta}  +  \frac{N-8}{2(N+2)(N+12)}\begin{bmatrix}
     \mathbbm 1_{N} & \\
    & - \frac{N}{12} \mathbbm 1_{12}
\end{bmatrix}\;,
\ee
where the second term acts only on the $\eta$ fields, and 
\be
\check{Q}= Q_{\psi\eta}  +  \frac{N-8}{2(N+2)(N+12)}\begin{bmatrix}
     1 & \\
    & - 1
\end{bmatrix}\;.\\
\ee
with the second term acting only on $\psi$'s.  $Q_{\psi\eta}$  is the generator of the anomaly-free combination of  $U_{\psi}(1)$ and  $U_{\eta}(1)$, with charges, $\frac{1}{2(N+2)}$ for $\psi$ and   $-\frac{1}{N+12}$ for $\eta$, respectively. 
%

 By assuming the candidate massless baryons:
 \be   B^{[CD], 1}=     \psi^{\{ij\},1}  \eta_{i}^C \eta_{j}^D \,, \label{onlythis} 
\ee
\be
  {\hat  B}^{A_1 B_1}=      \psi^{\{ij\},\, 2}   \{\overline {\psi^{\{ki\,, 1\}}   \eta_k^{A_1} } \}
\{\overline     {\psi^{\{\ell j\,, 1\}}   \eta_{\ell}^{B_1} }    \} \;,
\ee
and 
\be   {\tilde B}^{A_1,A_2}=   \chi_{ij}   \{\psi^{\{ik\},\, 1} \eta_{k}^{A_1} \} \{ \psi^{\{i\ell\},\, 1} \eta_{\ell}^{A_2}\} \;,
\ee
the natural anomaly matching is seen to work perfectly well,  see  Table~\ref{decomposition21} \footnote{The problem noted in 
\cite{BK}  was due to an erroneous assumption that all massless baryons should be composites made of three fermions, i.e., only of the type,  (\ref{onlythis}).}.

  \begin{table}[h!t]
  \centering 
  \begin{tabular}{|c|c|c|c|c|c|c|}
\hline
$ \phantom{{{   {  {\yng(1)}}}}}\!  \! \! \! \! \!\!\!$   & fields   &  $SU(N)_{\rm cf} $      &  $SU(12)$   &  $ {\tilde U}(1)   $  & $\hat{U}(1)$   &$\check{U}(1)$\\
 \hline
   \phantom{\huge i}$ \! \!\!\!\!$  {\rm UV} &  $\psi^{\{i  j\}, 1} $    &     $\Yvcentermath1  { { \yng(2)}} $     &   $ \frac{N(N+1)}{2} \cdot (\cdot)$    & $ 1 $  & $\frac{1}{2( N+2)}$ & $\frac{1}{N+12}$\\
   &  $\psi^{\{i  j\}, 2} $    &     $\Yvcentermath1 { { \yng(2)}} $     &   $ \frac{N(N+1)}{2}  \cdot (\cdot)$    & $  1 $ & $\frac{1}{2( N+2)}$ & $\frac{10}{(N+2)(N+12)}$\\
  & $\chi_{[i  j]} $   &   $\Yvcentermath1  {\bar  { \yng(1,1)}} $    &   $  \frac{N(N-1)}{2} \cdot  (\cdot) $    & $  -\frac{N-8}{N-2}$ & $0$    & $0$\\
 &  $  {\eta}^{B_1}_i  $      &  $\Yvcentermath1    {\bar  {\yng(1,1)}}\oplus   {\bar  {\yng(2)} }  $          &    $   N^2 \cdot (\cdot) $       &    $ 
  -1 $  & $-\frac{1}{2( N+2)}$& $-\frac{1}{N+12}$\\
 &  $  {\eta}^{B_3}_i    $      &  $   12   \cdot \Yvcentermath1    {\bar {\yng(1)}}  $         &    $N\cdot \Yvcentermath1 {\yng(1)} $       &    $-1$  & $-\frac{N+4}{24 (N+2)}$ & $-\frac{1}{N+12}$\\
  \hline 
 $ \phantom{{\bar{ \bar  {\bar  {\yng(1,1)}}}}}\!  \! \!\! \! \!  \!\!\!$  {\rm IR}  & 
  $ {\hat B}^{[A_1 B_1]}$         &   $ \Yvcentermath1  {\yng(2)}     $     &     $\frac{N(N+1)}{2}  \cdot (\cdot)$         &    $ 1 $  & $\frac{1}{2(N + 2 )}$ & $\frac{10}{(N+2)(N+12)}$\\[1.5ex]
   &   $ {\tilde B}^{[A_1 B_1]}$      &   $\Yvcentermath1 {\bar  {\yng(1,1)}} $          &    $   \frac{N(N-1)}{2}  \cdot  (\cdot) $       &    $  -\frac{N-8}{N-2}$ & $0$    & $0$\\[1.7ex]
  &  $ B^{[A_1 B_1], 1}$      &   $\Yvcentermath1     {\bar  {\yng(1,1)}}   $  &    $     \frac{N(N-1)}{2}  \cdot  (\cdot) $       &    $ -1 $  & $-\frac{1}{2(N+2)}$ & $-\frac{1}{N+12}$\\[1.7ex]
  &   $ B^{[A_1 B_3], 1}$         &   $ 12   \cdot  \Yvcentermath1   {\bar {\yng(1)}}  $     &    $   N \cdot \Yvcentermath1 {\yng(1)}$       &    $ -1 $ & $-\frac{(N+4)}{24 (N + 2)}$ & $-\frac{1}{N+12}$\\
\hline
\end{tabular}  
  \caption{\footnotesize  Single CF locking (\ref{singleCF}) in the $(2,1)$ model.   
  $A_1, B_1$  stand for  the flavor  indices up to  $N$;   $B_3$  the last $12$.  
  The natural anomaly matching is manifest, after the pair   $ \{  \psi^{\{i  j\}, 1}$,    the symmetric part of ${\eta}^{B_1}_i  \}$  forms a massive Dirac fermion and decouple.  
    }\label{decomposition21}
\end{table}

\newpage

\subsection{$SU(5)$  with  $(N_{\psi},N_{\chi}) = (1,9)$  \label{su5} }

A  nontrivial new example of color-flavor locked dynamical Higgs phase can be found in the context of  a tensorial  chiral gauge theory.  The matter fermions are   $\psi^{\{ij\}}$ and $\chi_{[ij]}^c$,   ($c=1,2,\ldots, 9$)  in symmetric and (anti-)antisymmetric 2-index tensor representations,    
\begin{equation}\label{modelSU(5)}
\Yvcentermath1
\yng(2) \oplus    9  \, \bar{\yng(1,1)}\;. 
\end{equation}
The classical symmetry group is 
\begin{equation}\label{classical_flavor}
SU(9) \times U(1)_{\psi} \times U(1)_{\chi}\;,  
\end{equation}
with a
 non-anomalous combination of the two $U(1)$ groups being
\begin{equation}
U(1)_{\psi \chi}: \qquad  \psi \to e^{27i \alpha}\psi, \qquad   \chi \to e^{-7i\alpha}\chi\;,  \qquad   \alpha \in [0, 2\pi)    \label{Upc1} 
\end{equation}

The implication of the presence of higher-form symmetries and associated generalized 't Hooft anomalies, has been 
studied systematically in the context of large classes of chiral gauge theories \cite{BKL1}-\cite{BKLproc}.  Only recently these analyses
have been extended to tensorial models such as this one \cite{AnberChan,BKLO}.

It turns out that the action of  the $\mathbb{Z}_5$ center of the color $SU(5)$ symmetry on the fermions can be compensated by the 
$U(1)_{\psi \chi}$  transformation with  
\begin{equation}\label{reproducing_Z5}
\alpha = \frac{2 \pi}{5}k\, ; \qquad k=1, \ldots, 5 \,,
\end{equation}
showing that the symmetry of the model is actually,   
\begin{equation}\label{symmSU(5)}
SU(5)_c \times \frac{SU(9) \times U(1)_{\psi \chi}}{\mathbb{Z}_5 \times  \mathbb{Z}_9  } \,;
\end{equation}
it  also possesses a color-flavor locked 1-form  $\mathbb{Z}_5$ symmetry \footnote{An analogue of the $\mathbb{Z}_N$ center symmetry in the pure $SU(N)$ Yang-Mills theory, which can be used,  by considering the Polyakov loop in the Euclidean formulation,  as a criterion of confinement / Higgs phase.
Analogously,  $\mathbb{Z}_9$ is the diagonal center symmetry  in $SU(9) \times U(1)_{\psi \chi} $. }

The consequences of (anomalies in) {\it   gauging this 1-form center symmetry}  can be  analyzed by a now well established fashion. 
Among others, one finds that $U(1)_{\psi \chi}$ transformation (\ref{Upc1}) induces an anomalous variation in the $4D$ effective action \footnote{This is sometimes called as a mixed  anomaly,   more specifically,  a mixed     $[U(1)_{\psi \chi}]-[\mathbb{Z}_5]^2$ anomaly.}, 
\begin{equation}\label{U1Z5}
\delta S^{4D}_{UV} = \frac{1}{8 \pi^2} 264375  \int_{\Sigma_4} \alpha (B^{(2)}_c)^2 = 10575 \, \alpha\, n\,;  \qquad n \in \mathbb{Z}_5\ne 0\;, 
\end{equation}
where  the the 1-form gauge field tensor $B^{(2)}_c$  carries a fractional 't Hooft flux, 
\begin{equation}\label{flux_Z_5}
\frac{1}{8 \pi^2} \int_{\Sigma_4} ( B^{(2)}_c)^2 = \frac{n}{25}\,, \qquad  n \in \mathbb{Z}_5\;.
\end{equation} 

The anomaly (the breaking) of $U(1)_{\psi \chi}$  found in the UV theory requires that  the IR  theory correctly reflects it.   Among other possibilities, here let us consider the possibility that  $U(1)_{\psi \chi}$   symmetry is broken by a bifermion condensate.  In the model  
(\ref{modelSU(5)}) the most attractive bifermion channel is  the one with two $\chi$  fermions forming the fundamental representation, 
\be  \Yvcentermath1      \bar{\yng(1,1)} \otimes  \bar{\yng(1,1)}  =   \yng(1) \oplus \ldots \;. 
\ee 
Let us assume that in the  vacuum the color $SU(5)$ is locked with the subgroup $SU(5)\subset SU(9)_{\chi}$:
\be     \epsilon^{ijk \ell m}  \brc  \chi_{jk}^A     \chi_{\ell m}^9 \ckt    \propto  \delta^{i  A}  \ne 0\;,  \qquad A=1,2,\ldots, 5\;,  \qquad 
\brc \psi\chi \ckt =0\;,  \label{condensBis} 
\ee
where the choice in the direction of the $SU(4)\subset SU(9)$ breaking, is arbitrary.  The symmetry  (\ref{symmSU(5)})  is broken to  
\be      SU(5)_{cf} \times SU(3) \times  U(1) \times U(1)^{\prime}\;,   \label{Symmetries}  
\ee
where  $U(1) \times U(1)^{\prime}$ are two $U(1)$ symmetries unbroken by the  condensate, (\ref{condensBis}).  The color-flavor locking 
(\ref{condensBis}) means that the first 5 flavor indices, $A=1,2,\ldots, 5$  of  $\chi_{jk}^A$   act as the antifundamental
${\bar 5}$ of the color-flavor locked $SU(5)$.     Let us then indicate (for simplicity of writing) the bifermion composite dynamical scalar  (\ref{condensBis})   as 
\be (\chi \chi)^i_A   \sim     \epsilon^{ijk \ell m}  \chi_{jk}^A     \chi_{\ell m}^9 \;.\ee
The massless baryons we expect in the infrared are 
\be    B^{\{A_1 A_2\}} =   \psi^{\{ij\}}  \overline {(\chi \chi)^i_{A_1} }  \overline {(\chi \chi)^j_{A_2} } \;;
\ee
\be {\hat B}_{[A_1 A_2], A} =     \chi_{[i  j]}^A   \,   (\chi \chi)^i_{A_1}   (\chi \chi)^j_{A_2}\;, 
\ee
(where the indices  $A, A_1, A_2$ run up to $5$,   forming  \be \Yvcentermath1  {\bar  { \yng(2,1)}}  \ee  of the flavor $SU(5)\subset SU(9)$);   
and 
\be {\hat B}_{[A_1 A_2], B} =     \chi_{[i  j]}^B   \,   (\chi \chi)^i_{A_1}   (\chi \chi)^j_{A_2}\;, 
\ee
where  the indices  $A_1, A_2$ run up to $5$,  $B=6,7,8$.

Note that,  \`a priori,    finding the solution of the anomaly matching requirement with respect to the unbroken symmetry group   (\ref{Symmetries})    (including its global structures, not explicitly shown there) with the known UV fermions  (the upper half of  Table~\ref{decomp19}),  
would have been a formidable task.   The solution, 
which can be  just read off Table~\ref{decomp19},   shows the power of the Natural Anomaly Matching mechanism.

  \begin{table}[h!t]
  \centering 
  \begin{tabular}{|c|c|c |c|c|c|  }
\hline
$ \phantom{{{   {  {\yng(1)}}}}}\!  \! \! \! \! \!\!\!$   & fields   &  $SU(5)_{\rm cf} $      &  $SU(3)$   &  $ U(1)   $      &   $ U(1)^{\prime}$  \\
 \hline
   \phantom{\huge i}$ \! \!\!\!\!$  {\rm UV} &  $\psi^{\{i  j\}} $    &     $ \Yvcentermath1 { { \yng(2)}} $     &   $ 15 \cdot (\cdot)$    & $0$    &   $3$   \\[1.5ex]
  & $\chi_{[i  j]}^A$   &   $\Yvcentermath1 {\bar  { \yng(2,1)}}  \oplus  {\bar  { \yng(1,1,1)}} $    &   $ 50  \cdot  (\cdot) $    & $  -3$   &  $-7$   \\[1.5ex]
  & $\chi_{[i  j]}^B$   &   $ 3 \cdot  \Yvcentermath1 {\bar  { \yng(1,1)}} $    &   $  10  \cdot   \Yvcentermath1 {\yng(1)} $    & $  4$   &  $7$   \\[1.5ex]
  & $\chi_{[i  j]}^9$   &   $\Yvcentermath1  {\bar  { \yng(1,1)}} $    &   $  10  \cdot  (\cdot) $    & $  3$   &   $7$    \\[1.5ex]
    \hline 
 $ \phantom{{\bar{ \bar  {\bar  {\yng(1,1)}}}}}\!  \! \!\! \! \!  \!\!\!$  {\rm IR}  &   $ B^{\{A_1 A_2\}} $      &   $\Yvcentermath1 { { \yng(2)}} $     &    $    15 \cdot  (\cdot) $       &    $ 0 $    &   $3$     \\[1.5ex]
  &    ${\hat B}_{[A_1 A_2], A}$        &   $ \Yvcentermath1 {\bar  { \yng(2,1)}}  $     &    $   40  \cdot  (\cdot) $       &    $ -3 $    &     $-7$   \\[1.5ex]
 &    ${\hat B}_{[A_1 A_2], B}$      &    $ 3 \cdot \Yvcentermath1 {\bar  {\yng(1,1)}} $      &    $  10  \cdot \Yvcentermath1 {\yng(1)}  $    &    $4$  &    $7$  \\[1.5ex]
\hline
\end{tabular}  
  \caption{\footnotesize  CF locking (\ref{condensBis}) and  the natural anomaly matching  in the $(1,9)$ model.   
  The index $A$  (and $A_{1,2}$) run up to $5$;     the index $B$ runs over  $B=6,7,8$. 
  The totally antisymmetric parts of  $\chi_{[i  j]}^A$  and $\chi_{[i  j]}^9$  pair up to condense,  form a massive Dirac fermion, and decouple.  
  The rest of the UV fermions are completely matched by the three types of baryons in the IR, in their quantum numbers.  Any conventional
  as well as  generalized anomaly matching is automatic. 
  }
  \label{decomp19}
\end{table}

\section{Infrared-free low-energy effective gauge theories   \label{IRfree} }     

In the previous section  systems in color-flavor-locked dynamical Higgs phase have been studied, where  the natural anomaly matching 
is seen at work. Another possibility is that the system is dynamically reduced to a low-energy  effective gauge theory which is infrared
free, either fully Abelian,  or with some nonAbelian gauge subgroup.   In  both these cases, the massless fermions in the low-energy effective theory are the subset of the UV fermions, which do not participate in the condensates and 
remain massless.  They become weakly coupled at low energies.   
We first briefly review the $\psi\chi\eta$ model  ($(N_{\psi}, N_{\chi})=(1,1)$) as an example of the systems which dynamically Abelianize,   and subsequently  discuss the possibility of a more general, infrared-free  low energy action characterized by a nonAbelian gauge group, in the context of the  model,  $(N_{\psi}, N_{\chi})=(1,9).$

\subsection{Dynamical Abelianization   \label{DA}} 

The $\psi\chi\eta$ model (the $(1,1)$  model in the classification  (\ref{classification}))   is  an $SU(N)$ theory with fermions 
\be   \psi^{\{ij\}}\;, \qquad  \chi_{[ij]}\;, \qquad    \eta_i^A\;,\qquad  A=1,2,\ldots 8\;,  
\ee
in the representation, 
\be  \Yvcentermath1   \yng(2) \oplus  {\bar  {\yng(1,1)}}  \oplus   8  \,  {\bar  {\yng(1)}}\;.   \label{fermions2}
\ee
The symmetry of the $\psi \chi \eta$ model is found to be \footnote{${\mathbbm Z}_N$  division takes care of the 
fact that   ${\mathbbm Z}_N =  SU(N) \cap  {\tilde U}(1)$, whereas  $ {\mathbbm Z}_{8/ N^*} $  arises from the intersection of     $U(1)_{\psi\chi} \times  {\tilde U}(1)\times  SU(8)$, see Eq.~(2.10)  of \cite{BKLDA}. 
}  
\be   G= \frac{ SU(N) \times  U(1)_{\psi\chi} \times  {\tilde U}(1)\times  SU(8)}{   {\mathbbm Z}_N    \times  {\mathbbm Z}_{8/ N^*} } \;.  \label{symmetry}
\ee
The two anomaly-free  $U(1)$ groups can be taken as 
\be   \tilde U(1): \qquad   \psi\to  e^{2i\alpha} \psi\;, \quad \chi \to   e^{ -2i\alpha} \chi\;, \quad \eta \to   e^{ -i\alpha} \eta\;, \label{def:tildeU1} \ee
and
\be U(1)_{\psi\chi}: \qquad   \psi\to  e^{i  \frac{N-2}{N^*} \beta} \psi\;, \qquad \chi \to   e^{- i \frac{N+2}{N^*}\beta} \chi\;,   \qquad \eta \to   \eta \;, \label{def:U1psichi}\ee  
where 
\be N^*=GCD(N+2, N-2)\quad \text{and}\quad  \alpha,\;\beta\in (0, 2\pi)\;.\ee

The system is asymptotically free, the first coefficient of the beta function being,
\be   b_0  = \frac 13\left[ 11N-  (N+2) -(  N-2) -8  \right] =  \frac{ 9 N- 8 }{3}\;.      \label{beta0}
\ee
Such a $\beta$ function suggests that the system is very strongly coupled in the infrared:   we certainly expect that something 
nontrivial occurs dynamically. To understand the fate of the symmetries  in this model  has long remained a puzzle \footnote{
The option that the system confines, with no global symmetry breaking and with some massless ``baryons''  saturating the 't Hooft anomalies, was judged not plausible \cite{GoityPecceiZ,Eichten:1985fs}.
Recent efforts to understand this system  \cite{BKS,BK} have eventually 
led to the  idea of dynamical Abelianization \cite{BKLDA, SheuShif}.}.

It  was proposed that a bifermion condensate 
  in the adjoint representation
\be   
\langle  \psi^{ik} \chi_{kj}   \rangle  = \Lambda^3   \left(\begin{array}{ccc}
      c_1  &  &  \\     & \ddots   &   \\   &  &    c_{N}
  \end{array}\right)^i_j   \;, \qquad  \brc  \psi^{ij}  \eta_j^A  \ckt    = 0
  \;,   \label{psichicond}  
\ee
\be         c_n    \in {\mathbbm C}\;,     \qquad     \sum_n c_n =0\;, \qquad       c_m - c_n  \ne 0\;, \ \     m \ne n  \;,  \label{psichicondBis}  
\ee
(with no other particular  relations among $c_j$'s)   forms,  inducing dynamical Abelianization of the color gauge group
\be    
SU(N)    \to    U(1)^{N-1} \;.      \label{SUNbreaking}    \ee
 The full symmetry is broken as
 \beq  SU(N) \times  \frac{  SU(8)_{\rm f} \times {\tilde U}(1)  \times  U(1)_{\psi\chi} }{   {\mathbbm Z}_N    \times  {\mathbbm Z}_{8/ N^*}  }    \longrightarrow    \frac{\prod_{\ell=1}^{N-1}  U(1)_{\ell}  \times SU(8)_{\rm f} \times {\tilde U}(1) }{ \prod_{\ell=1}^{N-1} {\mathbbm Z}_\ell \times {\mathbbm Z}_N \times {\mathbbm Z_2}  }  \;,
\label{fullsymbreaking}
\eeq
where
\be    {\mathbbm Z}_N  = U(1)_{N-1} \cap  {\tilde U}(1) =   SU(N) \cap  {\tilde U}(1)  \;.
\ee
For details see \cite{BKLDA}.     The anomaly matching can be read off from Table~\ref{SimplestDA}.  
 \begin{table}[h!t]
  \centering 
  \begin{tabular}{|c|c|c |c| }
\hline
$ \phantom{{{   {  {\yng(1)}}}}}\!  \! \! \! \! \!\!\!$   & fields      &  $ SU(8)$     &   $ {\tilde U}(1)   $  \\
 \hline
  \phantom{\huge i}$ \! \!\!\!\!$  {\rm UV}& $\psi$      &    $   \frac{N(N+1)}{2} \cdot  (\cdot) $    & $  \frac{N(N+1)}{2} \cdot (2)$    \\
  &  $\chi$      &    $   \frac{N(N-1)}{2} \cdot  (\cdot) $    & $  \frac{N(N-1)}{2} \cdot (-2)$        \\
 &$ \eta^{A}$      &   $ N  \cdot \Yvcentermath1 {\yng(1)}  $     &   $  8N \cdot  (-1) $ \\[0.5ex]
   \hline 
  \phantom{\huge i}$ \! \!\!\!\!$  {\rm IR}&       $ \psi^{ii}  $      &  $ N \cdot ( \cdot)   $        &    $  N \cdot (2) $   \\
     &  $ \eta^{A} $      &  $ N \cdot \Yvcentermath1 {\yng(1)}  $        &    $  8  N \cdot (-1) $   \\[0.5ex]
\hline
\end{tabular}  
  \caption{\footnotesize  Full dynamical Abelianization in the $\psi\chi\eta$ model.   The nondiagonal components of $\psi^{\{ij\}}$  and  $\chi_{[ij]}$  fields  pair up to form the condensates, (\ref{psichicond});  all the remaining fermions remain massless and become the infared
  degrees of freedom.}
\label{SimplestDA}
\end{table}
The massless fermions  in the infrared are shown again in Table~\ref{SimplestDA2},   where their quantum numbers with respect to the weak 
 $\prod_{\ell=1}^{N-1}  U(1)_{\ell}  \subset SU(N)$ are also shown.  
\begin{table}[h!t]
  \centering 
  \begin{tabular}{|c|c|c|c|c|c|c|c|c| }
\hline
$ \phantom{{{   {  {\yng(1)}}}}} \!  \! \! \! \! \!\!\!$   & fields    & $U(1)_1$ & $U(1)_2$  &   $\cdots $  &  $U(1)_{N-1}$    &  $ SU(8)$     &   $ {\tilde U}(1)   $    &   $ U(1)_{\rm an} $      \\
 \hline
   $\psi$  $ \phantom{{{   {  {\yng(1)}}}}}\!  \! \! \! \! \!\!\!$   & $\psi^{11}$   &   $1$     &   $\frac{1}{\sqrt{3}}$        &  $\cdots$   &  $\frac{2}{\sqrt{2N(N-1)}}$    &    $  (\cdot) $       &    $ 2$     & $ 1$      \\
                                                              &  $\psi^{22}$  &   $-1$   &   $\frac{1}{\sqrt{3}}$       &  $\cdots$  &   $\frac{2}{\sqrt{2N(N-1)}}$    &     $  (\cdot) $       &   $ 2$   & $ 1$   \\
                                                              &  $\psi^{33}$  &   $0$    &   $-\frac{2}{\sqrt{3}}$   &  $\cdots$  &   $\frac{2}{\sqrt{2N(N-1)}}$   &     $  (\cdot) $       &   $ 2$   & $ 1$   \\
                                                              &  $\vdots$       &             &                   &                  &      $\vdots$      &       $\vdots$       &    $\vdots$     & $\vdots$     \\
                                                              &  $\psi^{NN}$  &   $0$   &   $0$          &  $\cdots$      &   $-  \frac{2(N-1)}{\sqrt{2N(N-1)}}$     &     $  (\cdot) $       &          $ 2$    & $ 1$       \\
    \hline 
  $\eta$  $ \phantom{{{   {  \bar{\yng(1)}}}}}\!  \! \! \! \! \!\!\!$ &   $ \eta^a_1$  &    $ -\frac{1}{2}$       &    $-\frac{1}{2\sqrt{3}}$          &  $\cdots$  &      $-\frac{1}{\sqrt{2N(N-1)}}$    &       $\Yvcentermath1 \yng(1) $   &    $  -1  $      & $0$    \\
                                 &   $ \eta^a_2$  &  $\frac{1}{2}$       &    $-\frac{1}{2\sqrt{3}}$          &  $\cdots$  &      $-\frac{1}{\sqrt{2N(N-1)}}$   &       $\Yvcentermath1 \yng(1) $   &    $  -1  $    & $0$   \\
                                 &   $ \eta^a_3$  &  $0$                      &    $\frac{1}{\sqrt{3}} $          &  $\cdots$  &     $-\frac{1}{\sqrt{2N(N-1)}}$   &       $\Yvcentermath1 \yng(1) $   &    $  -1  $    & $0$     \\ 
                                                     &  $\vdots$         &           &                   &                &                    &       $\vdots$     &  $\vdots$     &  $\vdots$       \\
                                                             &   $ \eta^a_N$  &   $ 0 $      &    $0$          &  $\cdots$  &     $\frac{N-1}{\sqrt{2N(N-1)}}$   &       $\Yvcentermath1 \yng(1) $   &    $  -1  $     &    $0$     \\
                                                                 \hline 
   $\pi$  $ \phantom{{{   {  {\yng(1)}}}}}\!  \! \! \! \! \!\!\!$  &   ${\tilde \phi} \sim  (\psi\chi)^1_1$  & $0$  &    $0$      &    $\cdots$   &   $0$             &      $ (\cdot) $      &    $  0  $  &  $0$    \\
\hline
\end{tabular}  
  \caption{\footnotesize Massless fermions in the infrared in the $\psi\chi\eta$ model and their charges with respect to the unbroken symmetry groups.  The massless NG boson
  carries  no charges with respect to the unbroken symmetries.
  The two  (nonanomalous and anomalous)   $U(1)$ symmetries
  $ {\tilde U}(1)$    and    $ U(1)_{\rm an}   $
   which are not affected by the    $\psi\chi$  condensate are defined in  (\ref{def:tildeU1}) and in  Eq.~(2.7) of \cite{BKLDA}.  
 }
\label{SimplestDA2}
\end{table}

These assumptions have been carefully examined against the possible generalized anomalies 
\cite{GKSW}-\cite{WanWang}
arising from the gauging of the color flavor locked ${\mathbbm Z}_N$  symmetry shown in (\ref{symmetry}) and (\ref{def:tildeU1}), 
in \cite{BKLDA}.    Most significantly,  the latter gives rise to an anomalous variation  of the $4D$ effective action  
 \be   -  \frac{4 N^2}{N^*}   \delta  A_{\psi\chi}^{(0)} \,  \frac{1}{8\pi^2} \int_{\Sigma_4}    (B_{\rm c}^{(2)})^2\;,         \label{however}
  \ee
where $ \delta  A_{\psi\chi}^{(0)} =\beta $   is the  $U_{\psi\chi}(1)$ variation (\ref{def:U1psichi})  and  the  2-form   $B_{\rm c}^{(2)}$  fields 
introduced to  gauge the 1-form CF locked  ${\mathbbm Z}_N$  symmetry  carry 
a fractional   't Hooft fluxes
      \be      \frac{1}{8\pi^2}  \int_{\Sigma_4}    (B^{(2)}_\rmc)^2  \in   \frac {{\mathbbm Z}}{N^2}\;.  \label{fraction} 
      \ee
      Therefore  $U_{\psi\chi}(1)$ symmetry becomes anomalous, whereas  the  other  $U(1)$ symmetry, ${\tilde U}(1)$ (see (\ref{def:tildeU1})), as well as  $SU(8)$,   are not affected by the gauging of the   CF locked  ${\mathbbm Z}_N$. 
      These  results appear to fully support the assumption of the dynamical Abelianization of the system \cite{BKLDA}.

\subsection{Infrared-free nonAbelian color gauge group}

The $SU(5)$  tensorial gauge theory with fermions
 $\psi^{\{ij\}}$ and $\chi_{[ij]}^c$,   ($c=1,2,\ldots, 9$)  in  representations,    
\begin{equation}\label{modelSU(5)Bis}
\Yvcentermath1
\yng(2) \oplus    9  \hspace{1mm} \bar{\yng(1,1)}\;. 
\end{equation}
with classical symmetries 
\begin{equation}\label{classical_flavor}
SU(9) \times U(1)_{\psi} \times U(1)_{\chi}\;,  
\end{equation}
has been discussed in Sec.~\ref{su5},  to illustrate a nontrivial example of color-flavor locking 
(\ref{condensBis}),  the consequent dynamical Higgs phase, and Natural Anomaly Matching.
   
It is possible, instead, that the system goes through dynamical Abelianization, as in the $\psi\chi\eta$   model of the previous section, via condensate of the form, 
\begin{equation}\label{condensateDiag}
\ev{\psi^{ik}\chi^A}_{kj}  \propto \begin{bmatrix}
c_1 &  &  & \\
 & c_2 & & \\
 & &  \ddots \\
 & & & c_N
\end{bmatrix}^i_j\;, \qquad  \brc \chi\chi\ckt =0\;,  
\end{equation}
such that
the gauge symmetry is broken as   ($N=5$) 
\begin{equation}\label{dyn_ab_gen}
SU(N) \rightarrow U(1)^{N-1}\;. 
\end{equation}
The analysis in this case is similar to the one made in Sec~\ref{DA}.

Still another possibility, recently proposed in \cite{BKLproc},  is that the condensate of the  composite $\psi\chi$ field in the adjoint representation  forms, in the diagonal form (\ref{condensateDiag}),  but the  pattern of the gauge symmetry breaking is
\begin{equation}\label{NAbreaking1}
G_C= SU(5) \rightarrow     {\tilde G}_c  =    SU(2)_1 \times SU(2)_2 \times U(1)_1 \times U(1)_2\;.   
\end{equation}
The two unbroken $U(1)$  subgroups of $SU(5)$ may be chosen as 
\begin{equation}\label{U(1)1_gen}
U(1)_1:  \quad 
\begin{bmatrix}
2 \textbf{1}_{2\times 2} &  &  \\
& -\textbf{1}_{2 \times 2} &  \\
& & -2
\end{bmatrix} \;, \qquad    
U(1)_2:  \quad 
\begin{bmatrix}
\textbf{1}_{2\times 2} & & \\
& -4\textbf{1}_{2 \times 2} &  \\
& & 6  
\end{bmatrix}\;. 
\end{equation}

From the quantum numbers of the fermions which remain massless (see Table~\ref{tableIR}  below) it is easy to check that both $SU(2)\subset SU(5)$ (as well as the $U(1)$) factors are infrared free.  Assuming $A=9$ in (\ref{condensateDiag}),  the flavor  symmetry is broken as
\begin{equation}  \label{NAbreaking2}
G_f=  SU(9) \times U(1)_{\psi \chi} \rightarrow     {\tilde G}_f=    SU(8) \times U(1)'
\end{equation}
where 
$U(1)' \subset U(1)_{\psi \chi} \times SU(9)$.
The fermions in the UV are decomposed as the direct sum of representations in the unbroken color and flavor group.

We are left with the fermions in the IR shown in Table \ref{tableIR}, that reproduce 
 all  conventional 't Hooft anomalies, as well as  the generalized anomalies taking into account their global structure,   associated with  the unbroken symmetries  $  {\tilde G}_c \times {\tilde G}_f$ ((\ref{NAbreaking1}),  (\ref{NAbreaking2})).

\begin{table}[H]
\begin{center}
\begin{tabular}{ccccccc}
\toprule
fields & $SU(2)_1$ & $SU(2)_2$ & $U(1)_1$ & $U(1)_2$ & $SU(8)$ & $U(1)'$\\
\midrule
$\psi^{\lbrace ij \rbrace}$ & \Yvcentermath1 $\yng(2)$ & $3 \hspace{1mm} (\cdot)$ & 4 & 2 & $3 \hspace{1mm} (\cdot)$ & 27\\[1.5ex]
$\psi^{\lbrace IJ \rbrace}$ & $3 \hspace{1mm} (\cdot)$ & \Yvcentermath1 $\yng(2)$ & -2 & -8 & $3 \hspace{1mm} (\cdot)$ & 27\\[1.5ex]
$\psi^{55}$ & $(\cdot)$ & $(\cdot)$ & -4 & 12 & $(\cdot)$ & 27\\
\midrule
$\chi^9_{\left[ ij\right] }$ & $(\cdot)$ & $(\cdot)$ & -4 & -2 & $(\cdot)$ & -27\\[1.5ex]
$\chi^9_{\left[ IJ \right] }$ & $(\cdot)$ & $(\cdot)$ & 2 & 8 & $(\cdot)$ & -27\\[1.5ex]
\midrule
$\chi^c_{\left[ ij\right] }$ & $8 \hspace{1mm} (\cdot)$ & $8 \hspace{1mm} (\cdot)$ & -4 & -2 & \Yvcentermath1 $\yng(1)$ & -9/2\\[1.5ex]
$\chi^c_{iJ}$ & \Yvcentermath1 $16 \hspace{1mm} \yng(1)$ & \Yvcentermath1 $16 \hspace{1mm} \yng(1)$ & -1 & 3 & \Yvcentermath1 $4 \hspace{1mm} \yng(1)$ & -9/2\\[1.5ex]
$\chi^c_{\left[ IJ \right] }$ & $8 \hspace{1mm} (\cdot)$ & $8 \hspace{1mm} (\cdot)$ & 2 & 8 & \Yvcentermath1 $\yng(1)$ & -9/2\\[1.5ex]
$\chi^c_{i5}$ & \Yvcentermath1 $8 \hspace{1mm} \yng(1)$ & $16 \hspace{1mm} (\cdot)$ & 0 & -7 & \Yvcentermath1 $2 \hspace{1mm} \yng(1)$ & -9/2\\[1.5ex]
$\chi^c_{I5}$ & $16 \hspace{1mm} (\cdot)$ & \Yvcentermath1 $8 \hspace{1mm} \yng(1)$ & 3 & -2 & \Yvcentermath1 $2 \hspace{1mm} \yng(1)$ & -9/2\\
\bottomrule
\end{tabular}
\caption{\footnotesize  Representations and charges of the matter fermions in the IR theory  with respect to the continuous gauge and flavor symmetries unbroken by the condensate. The color indices run as  $i,j=1,2$,   $I,J=3,4$.    The multiplicities are also indicated. The fields $\psi^{iJ}$-$\chi^9_{iJ}$, $\psi^{i5}$-$\chi^9_{i5}$ and $\psi^{I5}$-$\chi^9_{I5}$ form massive Dirac pairs and decoupled. }\label{tableIR} 
\end{center}
\end{table}

 \section{Conclusion \label{Concl} }

 In this note we revisited the idea of Natural Anomaly Matching, which was encountered earlier  in the study of chiral gauge theories,  
 and which indicated that the certain dynamical assumption being considered  (e.g., Higgs mechanism caused by color-flavor-locked bifermion condensates) was a consistent one. Here we have streamlined the logics underlining it  (Sec.~\ref{NAM}), 
 and reviewed a few simple known examples, correcting some earlier statements, and discussed a few new applications   (Sec.~\ref{CF}, Sec.~\ref{IRfree}). 
 Natural Anomaly Matching works when the strong gauge interactions cause certain bifermion condensate to form, which yields dynamical Higgs phase with color-flavor locking, or brings the system into low-energy infrared-free effective gauge theories.  
 
One of the  inputs from the recent development involving the generalized (e.g., 1-form) symmetries and the idea of mixed anomalies,  is the following. In many chiral gauge theories studied, these mixed anomalies  indicate  some sort of dynamical color gauge symmetry breaking.  The hypothesis of confining vacua with no flavor symmetry breaking and saturation of the anomalies with composite gauge-invariant fermions in the IR, is found to be inconsistent \cite{BKL2}-\cite{BKLproc}. 

 Even though in general there is no way to predict exactly which bifermion condensate forms, often it is possible to make an educated guess, with the help of MAC (the maximally attractive channel) criterion \cite{RabyDim}, or by relying on the strong-anomaly argument \cite{BKL5}.  The fact that once such a choice among few possibilities is made the anomaly-matching constraints (the proper realization of the symmetries in the infrared) are fully guaranteed,  including the new, generalized mixed anomalies and global anomalies,  shows that Natural Anomaly Matching is a rather powerful idea,  in spite of its simple appearance.

 \section*{Acknowledgments }
 
 The present work is supported by the INFN special initiative project GAST (Gauge and String Theories).  
 We thank Jarah Evslin for useful discussions.

\end{document}